\documentclass[twocolumn,aps,showpacs]{revtex4}
\usepackage{graphicx,psfrag}
\def\bc{\begin{center}}
\def\ec{\end{center}}

\def\beq{\begin{equation}}
\def\eeq{\end{equation}}

\begin{document}

\title{
Creation of ray modes by strong random scattering
}

\author{K. Ziegler}
\affiliation{Institut f\"ur Physik, Universit\"at Augsburg\\
D-86135 Augsburg, Germany\\
}

\begin{abstract}
In the presence of strong random scattering the behavior of particles with degenerate spectra is 
quite different from Anderson localization of particles in a single band: 
it creates geometric states rather than confining the particles to an
area of the size of the localization length. These states are subject to 
a Fokker-Planck dynamics with universal drift velocity and disorder dependent diffusion coefficient.
This behavior has some similarity with the unidirectionally propagating edge states in quantum Hall systems.
\end{abstract}


\maketitle


Random scattering of wave-like states (electrons, photons or acoustic waves) leads either 
to diffusion for weak random scattering or to Anderson localization (AL) in the presence of strong random 
scattering.  AL is a phenomenon where diffusion or propagation is suppressed because random scattering 
confines the modes to a finite region, whose size is characterized by the localization length 
\cite{anderson58,abrahams79,wegner79}. The characteristics of AL can be observed in a pure form for photons 
\cite{genack89,lagendijk97,maret13} due to the absence of an additional particle-particle interaction. 

Another interesting phenomenon in disordered systems is the quantum Hall effect,
which is characterized by plateaux in the Hall conductivity \cite{klitzing80}.
The latter have been attributed to propagating edge modes in the otherwise 
Anderson localized bulk system \cite{qhe0}. Edge modes also exist in gapped systems without
any random scattering. A simple case is a massive 2D Dirac Hamiltonian, whose mass sign 
changes by crossing an edge in $y$-direction: The mass is $-m$ for $x<0$ and $m$ for $x>0$. The gap $2|m|>0$
prevents the system to generate any other extended state than that along the edge.
This mode decays exponentially when we go away from the edge in $x$-direction. 
The appearence of such states can be realized in a photonic crystal with Faraday effect \cite{haldane08}.

It was recently observed that similar one-dimensional modes can also be generated spontaneously
by strong scattering in 2D systems without the existence of any edge \cite{1404,1412}. These states, 
which will be called ray modes subsequently, are created in systems with a generalized particle-hole 
symmetry. The latter implies spectral degeneracies \cite{1412}. This is caused by the fact that
the Hamiltonian $H$ acts on states $\Psi_{r j}$ that depend on space coordinates $r$ and on an
additional spinor index $j$, typically with values $j=1,2$. Then there exists a unitary matrix
$U$ that acts only on the spinor index and which transforms $H$ into $-H^*$ as 
\beq
H=-UH^*U^\dagger
\ .
\label{symm00}
\eeq

In this paper we will show that ray modes can be created spontaneously by strong random scattering
in systems with spectral degeneracies based on the relation (\ref{symm00}).
Starting point is the transition probability for a particle, governed by the random Hamiltonian $H$,
to move from the site $r'$ on a lattice to another lattice site $r$ within the time $\tau$: 
\beq
P_{rr'}(\tau)=\sum_{j,j'}\langle|\langle r,j|e^{-iH\tau}|r',j'\rangle|^2\rangle_d
\ ,
\label{trans_prob}
\eeq
where $\langle r,j|...|r',j'\rangle$ is the quantum average and
$\langle ... \rangle_d$ is the average with respect to randomly distributed disorder.
$r$ and $r'$ refer to real space coordinates and the indices $j,j'$ refer to different 
bands of the system.
$P_{rr'}(\tau)$ is a fundamental quantity from which we can obtain transport and localization properties
\cite{thouless74,stone81}.

In the following we will focus on 
2D Dirac and 3D Weyl particles and on particles on the square lattice with $\pi$ flux.
For all these models the Hamiltonian reads in sublattice representation
\beq
H= {\vec \sigma}\cdot{\vec H} + \sigma_0 V
\ ,
\label{hamiltonian}
\eeq
where $\sigma_j$ ($j=0,1,2,3$) are Pauli matrices with the $2\times2$ unit matrix $\sigma_0$
and a random potential $V$ with mean zero and variance $\langle V_r V_{r'}\rangle_d=g\delta_{rr'}$. 
The part with $\sigma_1$, $\sigma_2$ provides scattering between different values of $j$, 
which will be crucial for the subsequent discussion.

The relation (\ref{symm00}) is satisfied, for instance, for the block-diagonal
matrix of the 3D gapless Weyl Hamiltonian
$diag(\sigma_1p_1+\sigma_2p_2 + \sigma_3p_3 + V\sigma_0,\sigma_1p_1+\sigma_2p_2 - \sigma_3p_3-V\sigma_0)$
with
\beq
U=\frac{1}{\sqrt{2}}\pmatrix{
0 & \sigma_1 \cr
\sigma_1 & 0 \cr
}
\ ,
\label{U}
\eeq
where $p_j$ is the momentum operator with $p_j^*=-p_j$. Another example is the massive 2D Dirac Hamiltonian
$diag(\sigma_1p_1+\sigma_2p_2 + \sigma_3m + V\sigma_0,\sigma_1p_1+\sigma_2p_2 + \sigma_3m-V\sigma_0)$, 
which also obeys condition (\ref{symm00}). 

It was shown in Ref. \cite{1404} that the Fourier component ${\tilde P}_{rr'}(i\epsilon)$ of $P_{rr'}(\tau)$ 
agrees for large distances $|r-r'|$ with a correlation function of a random-phase model, defined
by the random matrix
\beq
C_{rr'}= 2\delta_{rr'}-\sum_{j,j'}h_{rj,r'j'}\sum_{j'',r''}h^\dagger_{r'j',r''j''}
\ ,
\eeq
where the propagator
\beq 
h_{rr'}=\sigma_0\delta_{rr'}+ 2i\eta({\cal H} - i{\bar\eta}\sigma_0)^{-1}_{rr'}
\label{def_h}
\eeq
depends on the random phase Hamiltonian
\beq
{\cal H}_{rj,r'j'}=e^{i\alpha_{rj}}{\bar H}_{rj,r'j'} e^{-i\alpha_{r'j'}}
\ .
\label{hamiltonian1}
\eeq
${\bar H}=\langle H\rangle_d$ is the average Hamiltonian and $\eta$ is the scattering rate
while ${\bar\eta}=\eta+\epsilon$.
$\eta$ can be considered as an empirical parameter or can be calculated self-consistently
from the self-energy of the average one-particle Green's function $\langle(H-z)^{-1}\rangle_d$ 
\cite{scba}. In any case, it increases with variance $g$ of the random potential.
  
In the limit $\epsilon\to 0$ the propagator $h$ is unitary since
\beq
hh^\dagger={\bf 1}-4\epsilon\eta({\cal H}^2+{\bar\eta}^2\sigma_0)^{-1}
\ .
\label{unitary0}
\eeq
Then there is the following asymptotic relation for large scales between the random phase 
model and the Fourier components of the average transition probability \cite{1404}:  
\beq
{\tilde P}_{rr'}(i\epsilon)\sim 
\frac{\langle C^{-1}_{rr'}\det C\rangle_a}{\langle\det C\rangle_a}
\ ,
\label{corr00}
\eeq
where the brackets $\langle ...\rangle_a$ mean integration with respect to the angular variables 
$\{0\le\alpha_{rj}<2\pi\}$, normalized by $2\pi$. These random angles represent the relevant
part of the disorder fluctuations in terms of long-range correlations. In other words, there
is a  mapping from the original random Hamiltonian in Eq. (\ref{hamiltonian}) 
to the random phase Hamiltonian ${\cal H}$ that preserves the long-range correlations.

The expression (\ref{corr00}) can be calculated for strong scattering in powers of the expansion
parameter $E_b/\eta$ ($E_b$: bandwidth), combined with a mean-field
approximation as the starting point for the expansion. It has to be chosen such that the
expansion is convergent.


{\it Mean-field approximation:}
The (unnormalized) distribution density $\det C$ in Eq. (\ref{corr00}) is approximated by a constant
phase $\alpha_{rj}\approx{\bar\alpha}_j$ and the exact value is approached systematically 
by a convergent expansion in terms of the phase factor fluctuations $e^{\alpha_{rj}}-e^{{\bar\alpha}_j}$.
First, it should be noticed that there is an invariance of $C$ with respect to a global phase change
$\alpha_{rj}\to \alpha_{rj}+\phi$. This implies that the mean-field distribution depends only
on the relative phase difference $\Delta={\bar\alpha}_1-{\bar\alpha}_2$. Its value is determined by the
condition $\max_{\Delta}\int_p\log(|{\tilde C}_p(\Delta)|)$, where ${\tilde C}_p(\Delta)$ are
the Fourier components of $C$ with uniform phases. This condition may have several solutions 
$\Delta_l$ with the same maximum, such that we have to sum over all of them:
\beq
\det C\approx\sum_l 
\exp\left[\int_p\log({\tilde C}_p(\Delta_l))\right]
\ .
\eeq
It turns out that the Fourier components ${\tilde C}_p(\Delta_l)$ can be written as
\beq
{\tilde C}_p(\Delta) =2-\pi_0-{\vec s}\cdot{\vec\pi}
\label{inv_prop}
\eeq
with $\pi_j=Tr_2(\sigma_j{\bar h}_p{\bar h}_{p=0}^\dagger)$ for $j=0,1,2$,
\beq
{\bar h}_p=\sigma_0+ 2i\eta({\bar H}_p - i{\bar\eta}\sigma_0)^{-1}
\ ,
\label{av_prop}
\eeq
and with the 2D unit vector ${\vec s}=(\cos\Delta,\sin\Delta)$.
Then Eq. (\ref{corr00}) reads in this mean-field approximation
\beq
{\tilde P}_{rr'}(i\epsilon)\approx 2\int_pe^{-ip\cdot(r-r')}\sum_l\frac{1}{{\tilde C}_p(\Delta_l)}
\ .
\eeq
The pole of $1/{\tilde C}_p(\Delta_l)$ with respect to $\epsilon$ gives, after the
analytic continuation $\epsilon\to i\omega$, the effective dispersion $\omega_p$ of the new mode.
This pole depends strongly on the details of the Hamiltonian components with Pauli matrices $\sigma_1$
and $\sigma_2$. It is important that we
are only interested in the long range regime of ${\tilde P}_{rr'}(i\epsilon)$; i.e., in the behavior
for small momenta. Therefore, we focus on ${\bar h}_p$ for small $p$, and according to the expressions
in (\ref{inv_prop}) and (\ref{av_prop}), this means that we need ${\vec s}\cdot{\vec H}_p$ 
for small $p$. In the following we calculate (\ref{inv_prop}) for three different models, namely
2D massive Dirac particles, 3D Weyl particles and particles on a square lattice with $\pi$ flux,
whose low energy Hamiltonians behave like $H\sim {\vec \sigma}\cdot{\vec p}$ for small $p$.

\begin{figure}[t]
\begin{center}
\includegraphics[width=5cm,height=3cm]{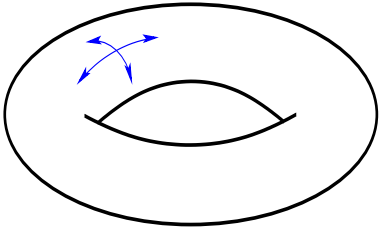}
\caption{Four ray modes on a torus, created by strong random scattering.
}
\label{fig:rays}
\end{center}
\end{figure}


{\it 2D Dirac particles:}
For massive 2D Dirac fermions on a torus the Hamiltonian reads
${\tilde H}_p={\vec\sigma}\cdot{\vec p}+\sigma_3m$ ($0\le p_j<2\pi$). Then we obtain
\[
{\tilde h}_p=\sigma_0+\frac{2i\eta}{p^2+m^2+{\bar\eta}^2}
\left({\vec\sigma}\cdot{\vec p}+\sigma_3m+i\sigma_0{\bar\eta}\right)
\]
and ${\vec s}=(\cos(\Delta+\theta), \sin(\Delta+\theta))$ for ${\tilde C}_p(\Delta)$,
which depends on the angle $\theta=\arctan(2m\eta/(m^2-{\bar\eta}^2))$. 
For $\epsilon\sim 0$ we obtain
\beq
\frac{1}{{\tilde C}_p(\Delta)}\sim\frac{(p^2+m^2+{\bar\eta}^2)/2\eta}
{2\epsilon+i{\vec s}\cdot{\vec p}+{\bar\eta}p^2/(m^2+{\bar\eta}^2)} 
\ ,
\label{corr_dirac}
\eeq
whose pole give the dispersion of a ray mode
\beq
\omega_p\sim \frac{1}{2}{\vec s}\cdot {\vec p}-iDp^2\ \ \ {\rm with}\ \
D=\frac{\eta}{2(m^2+\eta^2)}
\ .
\label{dispersion_dirac}
\eeq
The second term of $\omega_p$ is imaginary and describes damping. 
The mean-field condition $\max_{\Delta}\int_p\log(|{\tilde C}_p(\Delta)|)$
is solved by $\Delta+\theta=0,\pi/2,\pi,3\pi/2$. Thus, four rays are created by 
strong random scattering with unit vectors ${\vec s}_{1,3}=(\pm 1,0)$ and ${\vec s}_{2,4}=(0,\pm 1)$
(cf. Fig. {\ref{fig:rays}).
The linear dispersion without damping is shown on the right-hand side of Fig. \ref{fig:dispersions}.

The dispersion (\ref{dispersion_dirac}) can also be understood as a Fokker-Planck
dynamics with drift velocity ${\vec s}$ \cite{FPE}. 
This can be seen when we calculate the transition probability 
from the expression (\ref{corr_dirac}) within a Fourier representation in time and space. 
By assuming that the propagation starts from 
the initial site $0$ we get for the transition probability in Eq. (\ref{trans_prob}) for each
of the four ray modes
\beq
P_{{\vec r}}(\tau)\sim \frac{1}{\tau D4\pi}
\exp\left[-\frac{1}{4D\tau}({\vec r}+\tau {\vec s}_l)^2\right]
\eeq
with the effective diffusion coefficient $D$, which is the damping coefficient in Eq. (\ref{dispersion_dirac}).
This result describes diffusion with a propagating (drifting) center in the direction
of ${\vec s}_l$. The damping term in the dispersion (\ref{dispersion_dirac}) leads to an additional
diffusion away from these propagating centers. Diffusion slows down as the scattering
rate $\eta$ is increased. This indicates that strong random scattering confines (or localizes)
the movement of the particle to rays along the unit vectors ${\vec s}_l$. 

\begin{figure}[t]
\begin{center}
\includegraphics[width=3.5cm,height=5cm]{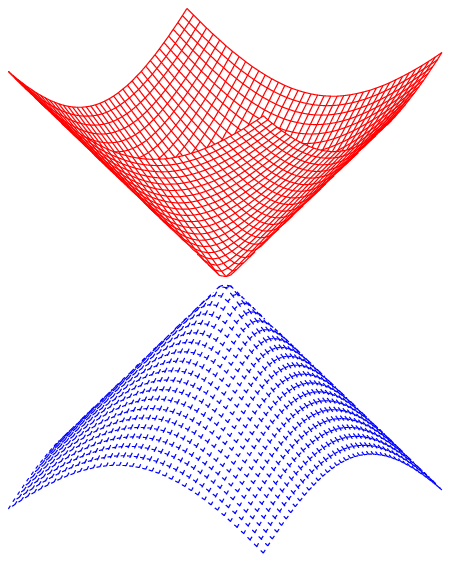}
\includegraphics[width=3.5cm,height=5cm]{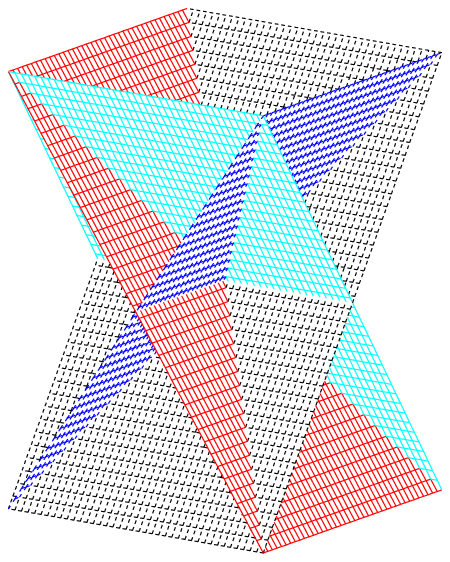}
\caption{Linear Dirac dispersion (left) and the linear dispersion of the four ray modes (right).
The latter consists of four intersecting planes whose slopes agree up to a factor 1/2 with
those of the Dirac dispersion.
}
\label{fig:dispersions}
\end{center}
\end{figure}


{\it 3D Weyl particles:}
The Hamiltonian for 3D Weyl particles with 3D momentum ${\vec p}$ reads ${\tilde H}_p={\vec p}\cdot {\vec \sigma}$.
This implies for the propagator
\[
{\tilde h}_p=\sigma_0+\frac{2i\eta}{p^2+{\bar\eta}^2}( {\vec\sigma}\cdot{\vec p}+i\sigma_0{\bar\eta})
\]
from which we obtain
\[
\frac{1}{{\tilde C}_p(\Delta)}
\sim \frac{(p^2+{\bar\eta}^2)/2{\bar\eta}}{2\epsilon+i{\vec s}\cdot{\vec p} +p^2/{\bar\eta}}
\ .
\]
Then the dispersion of the ray modes reads
\beq
\omega_p\sim \frac{1}{2}{\vec s}\cdot {\vec p}-iDp^2
\ .
\eeq
Since the mass is zero in this case, the damping coefficient is $D=1/2\eta$. 
Although the Hamiltonian is isotropic in 3D, there are again only four ray modes on the $x$-$y$ torus.


{\it Square lattice with $\pi$ flux:}
The square lattice is bipartite. Therefore, the tight-binding Hamiltonian for particles 
on a square lattice, which describes nearest-neighbor hopping matrix elements, can be written in
terms of two sublattices with coordinates $(rj)$ ($j=1,2$ and $r$ are the coordinates on
the sublattice with $j=1$). Including the Peierls phase factors for the $\pi$ flux
with $\pm i$ in $x$-direction and $\pm 1$ in $y$-direction, as indicated in Fig. 
\ref{fig:square}, the corresponding Hamiltonian reads
\beq
{\tilde H}_p={\vec\sigma}\cdot{\vec P} +\sigma_3m , \ \ \ P_j=t\sin a k_j
\ .
\label{square_ham}
\eeq
This is a discretized 2D Dirac Hamiltonian with a lattice constant $a$ and the hopping energy $t$.
The flux through a plaquette of the square lattice is $-1$, which is related to a phase $\pi$.
There is also a term with mass $m$ that represents a staggered potential with $m$ ($-m$) on 
sublattice $j=1$ (sublattice $j=2$). This Hamiltonian also obeys property (\ref{symm00}) as described for
the 2D Dirac Hamiltonian. The form of this Hamiltonian allows us to apply directly the results of the 
2D Dirac Hamiltonian that gives
\[
\frac{1}{{\tilde C}_p(\Delta)}\sim\frac{(P^2+m^2+{\bar\eta}^2)/2\eta}
{2\epsilon+i{\vec s}\cdot{\vec P}+{\bar\eta}P^2/(m^2+{\bar\eta}^2)} 
\ .
\]
This implies the dispersion
\beq
\omega_p\sim \frac{1}{2}{\vec s}\cdot {\vec P}-iDP^2
\eeq
for the ray mode on a square lattice with $\pi$ flux with the same directions
as in the two previous cases. Here it should be noticed that
there are small values for $P_j$ near the four spectral nodes with $ak_j=0, \pm\pi$. 

\begin{figure}[t]
\begin{center}
\includegraphics[width=4cm,height=4cm]{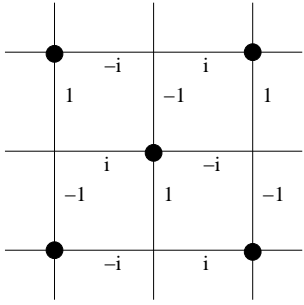}
\caption{Square lattice with $\pi$ flux as defined by the tight-binding Hamiltonian (\ref{square_ham}).
The dots represent the sublattice with index $j=1$ and the complex numbers are the Peierls
phase factors for hopping from $j=1$ to $j=2$. 
}
\label{fig:square}
\end{center}
\end{figure}


{\it Discussion:} The effect of strong random scattering in models with a Weyl-Dirac Hamiltonian 
and for the tight-binding Hamiltonian on the square lattice with $\pi$ flux does not lead
to conventional AL but creates four orthogonal ray modes. For periodic boundary conditions
the ray modes are parallel to a torus or follow its circumference (cf. Fig. \ref{fig:rays}). 
The spontaneous creation of the four ray modes can be understood as an azimuthal localization at four discrete angles.
In contrast to isotropic diffusion or isotropic Anderson localization, the states are confined to four orthogonal
directions. This is similar to the uni-directional edge states in the quantum Hall effect or in
photonic crystals with Faraday effect \cite{haldane08}. The mechanisms for their creation are rather different, though.
Whereas the edge states are created in gapped photonic crystals at interfaces between regions with
different Chern numbers \cite{haldane08}, strong randomness creates the ray modes spontaneously 
in an isotropic system with strong random scattering. 
This makes them more accessible in real materials, provided that a band structure with
linear dispersion exists for interband scattering.

The result of the calculation can be summarized as a mapping of the momentum part of the 
Dirac-Weyl Hamiltonian to the ray-mode dispersion as 
\beq
{\vec \sigma}\cdot{\vec p}\ \  \to \ \ \frac{1}{2}{\vec s}\cdot{\vec p} - iD p^2
\ .
\label{map}
\eeq
Thus, in all three cases we have considered here, the Pauli matrix vector ${\vec \sigma}$ 
is replaced by the vector $(1/2){\vec s}$,
whose components are scalars. Moreover, there is an additional damping term, whose coefficient
$D$ in Eq. (\ref{dispersion_dirac}) decreases with an increasing scattering rate and increasing gap. 
The mapping (\ref{map}) is reminiscent of the renormalization of the average one-particles Green's function
as found, for instance, in the form of a complex self-energy in the self-consistent Born approximation.
In the case of 2D Weyl particles we get from random scattering the mapping
\[
{\vec \sigma}\cdot{\vec p}\ \  \to \ \ \ {\vec \sigma}\cdot{\vec p}+i\sigma_0\eta
\ .
\]
In contrast to the constant scattering rate $\eta$, the damping of the ray modes vanishes 
quadratically for small momenta.

The fact that, for instance, photons can escape from a cloud of random scatterers is similar to
the phenomenon of Klein tunnelling in a system with a potential barrier \cite{klein29,katsnelson06}. 
As in the latter case, it is crucial that there are two spectral bands. In our calculation this
is reflected by the term ${\vec s}\cdot {\vec \pi}$ in (\ref{inv_prop}). It is sufficient for 
the generation of ray modes that this term vanishes linearly with the momentum.

Tight-binding models on the square or the honeycomb lattice without flux do not develop ray modes
because ${\vec s}\cdot{\vec \pi}$ (or equivalently ${\vec s}\cdot{\vec H}$) is not linear for
small momenta, as a calculation of the inverse propagator (\ref{inv_prop}) indicates. 
However, the linear behavior of ${\vec s}\cdot{\vec H}$ at small momenta might be obtained by 
breaking the time-reversal invariance: either by a periodic magnetic flux \cite{haldane88}, 
by an additional spin texture on the honeycomb lattice \cite{hill11},
or by the Faraday effect of a magneto-optic medium in a photonic crystal \cite{haldane08}.

\end{document}